\documentclass[aps,prb,preprint,groupedaddress,showpacs,amsmath,amssymb]{revtex4}

\usepackage{graphicx}
\usepackage{dcolumn}
\usepackage{bm}

\begin{document}

\title{First order structural phase transition in CaFe$_2$As$_2$.}



\author{N. Ni}
\author{S. Nandi}
\author{A. Kreyssig}
\author{A. I. Goldman}
\author{E. D. Mun}
\author{S. L. Bud'ko}
\author{P. C. Canfield}

\affiliation{Ames Laboratory US DOE and Department of Physics and Astronomy, Iowa State University, Ames, Iowa 50011}

\date{\today}

\begin{abstract}
CaFe$_2$As$_2$ has been synthesized and found to form in the tetragonal, ThCr$_2$Si$_2$ structure with lattice parameters $a = 3.912(68)~\AA$ and $c = 11.667(45)~\AA$.  Upon cooling through 170 K, CaFe$_2$As$_2$ undergoes a first order, structural phase transition to a low temperature, orthorhombic phase with a $2 - 3$ K range of hysteresis and coexistence. This transition is clearly evident in microscopic, thermodynamic and transport measurements. CaFe$_2$As$_2$ is the third member of the AFe$_2$As$_2$ (A = Ba, Sr, Ca) family to exhibit such a dramatic phase transition and is a promising candidate for studies of doping induced superconductivity.
\end{abstract}


\maketitle

\section{Introduction}
The discovery of superconductivity with critical temperatures between 30 and 55 K in two families of layered iron arsenides \cite{1,2,3,4} has caused a groundswell of enthusiasm and activity amongst experimentalists and theorists alike.  Although the superconducting transition temperatures are somewhat lower, the potassium doped, AFe$_2$As$_2$ (A = Ba and Sr) family of intermetallic compounds are a particularly interesting system because (i) their lack of oxygen, (ii) their well known ThCr$_2$Si$_2$ structure type, and (iii) the ability to grow substantial sized single crystals.  \cite{5,6,7,8}

The parent compounds, BaFe$_2$As$_2$ and SrFe$_2$As$_2$ each manifest a structural phase transition from a high temperature tetragonal to a low temperature orthorhombic unit cell. \cite{5,8,9} This structural phase transition is extremely sensitive to chemical or structural perturbation and is suppressed either partially or completely by K-doping as well as slight additions of Sn. \cite{4,5,6,7,8}  The fact that a similar structural phase transition also appears in the parent RFeAsO compounds and is suppressed by F-doping \cite{1} strongly links the superconductivity of the doped iron arsenides to a proximity to this structural instability.  In BaFe$_2$As$_2$ and SrFe$_2$As$_2$ there are indications that the phase transition may be first order in nature, \cite{4,5,8,ad} but, to date, there has not been definitive evidence to support this designation.  In this work we report the synthesis of the isostructural compound: CaFe$_2$As$_2$, a previously unknown member of the ThCr$_2$Si$_2$ structure group, and present microscopic, thermodynamic and transport data that clearly indicate that the tetragonal to orthorhombic phase transition, also present in this compound, is first order in nature.

\section{Experimental Methods}
Single crystals of CaFe$_2$As$_2$ were grown out of a Sn flux, using conventional high temperature solution growth techniques. \cite{5,10}  Elemental Ca, Fe and As were added to Sn in the ratio of [CaFe$_2$As$_2$] : Sn  =  1 : 48 and placed in a 2 ml alumina crucible.  A second, catch, crucible containing silica wool was placed on top of the growth crucible and sealed in a silica ampoule under approximately $1/3$ atmosphere of argon gas.  It should be noted that the packing and assembly of the growth ampoule was performed in a glove box with a nitrogen atmosphere.  The sealed ampoule was placed in programmable furnace and heated to 850 $^\circ$C and cooled over 36 hours to 500 $^\circ$C.  Once the furnace reached 500 $^\circ$C the Sn was decanted from the plate-like CaFe$_2$As$_2$ crystals as well as a yet to be identified rod like, crystalline, second phase.

\begin{figure}
\begin{center}
\includegraphics[angle=0,width=120mm]{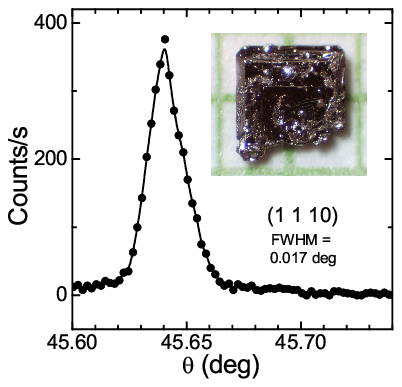}
\end{center}
\caption{(Color online) Rocking curve through the $(1~1~10)$ reflection of the CaFe$_2$As$_2$ single crystal used for the X-ray diffraction study.  Inset:  picture of a CaFe$_2$As$_2$ single crystal on mm grid paper.  The crystallographic $c$-axis is perpendicular to the plate of the crystal. The small droplets on the surface are residual Sn flux.}\label{F1}
\end{figure}

The inset to Fig. \ref{F1} shows a picture of a single crystal of CaFe$_2$As$_2$ against a mm scale.  Typical crystals have dimensions of $(1-2) \times (1-2) \times (0.1-0.2)$ mm$^3$.  Some crystals manifest linear dimensions as large as $3 - 4$ mm.  The crystallographic $c$-axis is perpendicular to the plane of the plate-like single crystals.  The BaFe$_2$As$_2$ and SrFe$_2$As$_2$ crystals are exceptionally micaceous in nature and rather brittle.  On the other hand, the CaFe$_2$As$_2$ single crystals are also layered and easy to cleave / separate, but instead of being brittle, they are somewhat malleable and tend to smear and shear when bent or when they are ground into a powder.

Powder X-ray diffraction data were collected on PANalytical  X'Pert Pro PW3040PRO MPD, with Cu $K \alpha$ radiation (45kV 40mA) selected by a graphite monochromator.  Single crystal X-ray diffraction measurements were performed on a standard four-circle diffractometer using Cu $K \alpha$ radiation from a rotating anode X-ray source, selected by a Ge$(1~1~1)$ monochromator. For these measurements, a plate-like single crystal with dimensions of $1.2 \times 1.2 \times 0.3$ mm$^3$ was selected. The sample was mounted on a flat copper sample holder on the cold-finger of a closed cycle displex cryogenic refrigerator with the $(0~0~1)-(1~1~0)$ reciprocal lattice plane coincident with the scattering plane. The diffraction patterns were recorded for temperatures between 10 K and 300 K and with the diffractometer optimized for high resolution for transverse scans in the scattering plane.  The measured mosaicity of this crystal was $0.017^\circ$ full width half maximum for the $(1~1~10)$ reflection at room temperature (Fig. \ref{F1}), indicating the excellent quality of the single crystal. This value is significantly better than for the AFe$_2$As$_2$ (A = Ba, Sr) crystals used in previous studies \cite{5,8}, and may be related to the absence, or strongly reduced content, of Sn flux incorporated into the structure.  Elemental analysis was performed (without standards) by EDS measurements in a JEOL model 5910v-SEM.

Magnetic field and temperature dependent magnetization data were collected using a Quantum Design (QD) Magnetic Properties Measurement System (MPMS) and temperature dependent specific heat as well as magnetic field and temperature dependent electrical transport data were collected using a QD Physical Properties Measurement System (PPMS).  Electrical contact was made to the samples using Epotek H20E silver epoxy to attach Pt wires in a 4-probe configuration.  (It is worth noting that curing of the epoxy at 120 $^\circ$C for up to 30 minutes does not seem to degrade the samples.)  Basal plane resistivity values were determined by estimating the length and cross sections of the samples assuming that the current flow was uniformly distributed throughout the cross section.  Given the layered nature and the thinness of the crystals we believe that the resistivity values should be accurate to better than $\pm 30\%$.  

\section{Results}
\begin{figure}
\begin{center}
\includegraphics[angle=0,width=120mm]{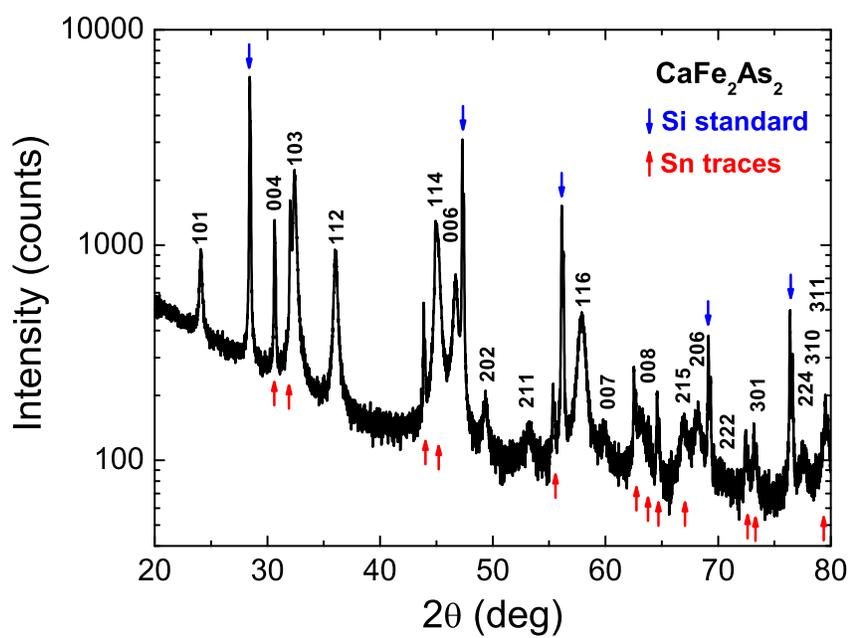}
\end{center}
\caption{(Color online) Powder X-ray diffraction spectrum of ground single crystal CaFe$_2$As$_2$.  Note that Si powder was added as a standard and Sn is present from residual flux on surface of samples used.}\label{F2}
\end{figure}

Figure \ref{F2} presents the powder X-ray diffraction spectra on ground single crystals of CaFe$_2$As$_2$.  Both the indexing of the observed reflections, as well as a room temperature Laue pattern taken on a single crystal, are consistent with the space group $I4/mmm$ (139), as reported for the high temperature phases of BaFe$_2$As$_2$ and SrFe$_2$As$_2$, with lattice constants $a = 3.912(68)~\AA$ and $c = 11.667(45)~\AA$.  The somewhat large uncertainty in the lattice parameters is associated with the somewhat larger than usual peak widths, which in turn are associated with the soft and ductile nature of the CaFe$_2$As$_2$ crystals.  As shown in Fig. \ref{F1} and as will be discussed below, the single crystal diffraction peaks were very sharp, indicating that as grown crystals were highly ordered.  Elemental analysis confirmed the Ca:Fe:As ratio to be $1:2:2$ within the instrumental error. Far more significantly, no Sn could be detected in the bulk of the CaFe$_2$As$_2$ sample by the EDS measurement.

\begin{figure}
\begin{center}
\includegraphics[angle=0,width=120mm]{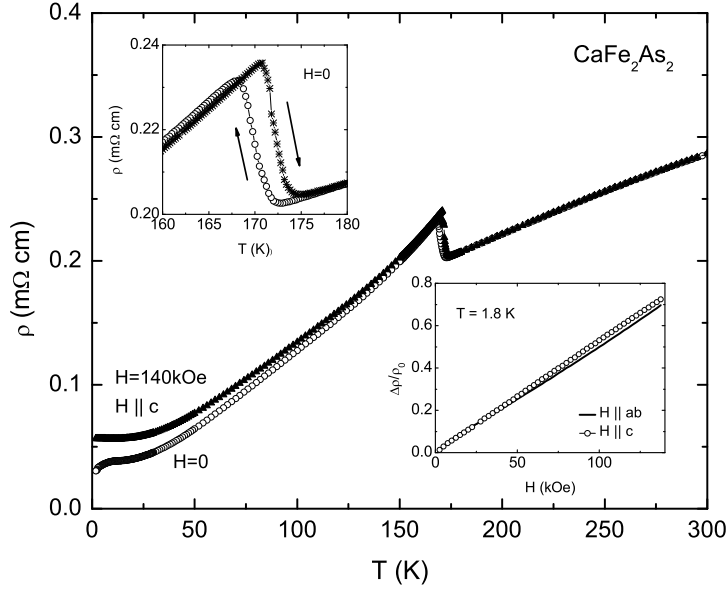}
\end{center}
\caption{Temperature dependent electrical resistivity of CaFe$_2$As$_2$ with current flowing within the basal plane for $H \| c$, $H = 0$ and 140 kOe.  Upper inset:  hysteresis in temperature dependent resistivity near the 170 K phase transition  (arrows indicate increasing and decreasing temperature scans).  Lower inset:  magnetoresistance for $H \| ab$ and $H \| c$ at $T = 2$ K. }\label{F3}
\end{figure}

Figure \ref{F3} presents the temperature dependent electrical resistivity of CaFe$_2$As$_2$.  The most conspicuous feature is the very sharp jump in the resistivity near 170 K.  The upper inset shows the hysteretic temperature dependence of this feature:  there is an $\sim 2$ K hysteresis between the cooling and warming curves.  This hysteresis is larger than the shift in the resistivity associated with the application of a 140 kOe magnetic field along the either $c$-axis or $ab$ plane.  There is a finite, low temperature magnetoresistance that (as shown in the lower inset) is essentially the same for both directions of applied field. It is worth noting that two features of the electrical transport data on CaFe$_2$As$_2$ set it apart from BaFe$_2$As$_2$ and SrFe$_2$As$_2$. \cite{5,8,9}  First, the very sharp increase in resistivity at $T \approx 170$ K is markedly different from the decrease in resistivity found in pure BaFe$_2$As$_2$ \cite{9} and SrFe$_2$As$_2$ \cite{8} as well as different from the more gradual increase found in Sn doped BaFe$_2$As$_2$. \cite{5} Secondly, the nearly isotropic, linear, low temperature  magnetoresistance found for CaFe$_2$As$_2$ is markedly different from the anisotropic and non-linear field dependence found in SrFe$_2$As$_2$. \cite{8} 

\begin{figure}
\begin{center}
\includegraphics[angle=0,width=120mm]{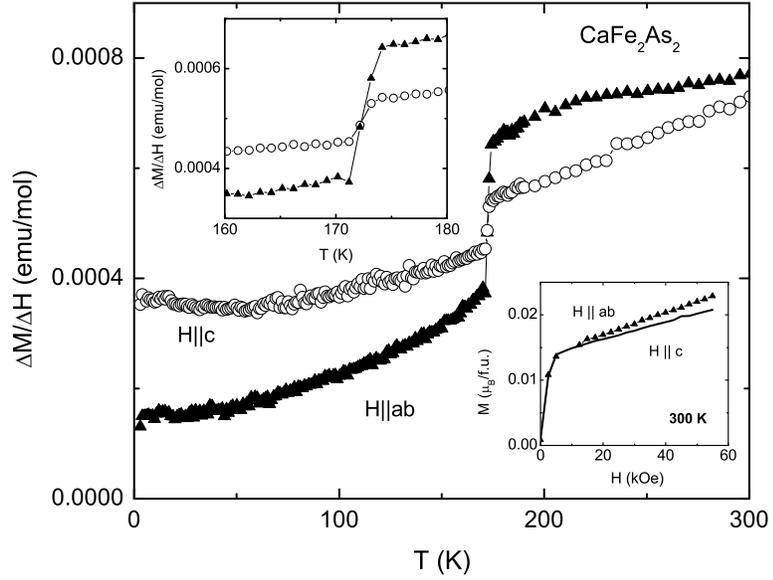}
\end{center}
\caption{Temperature dependent magnetic susceptibility of CaFe$_2$As$_2$ for applied field parallel to and perpendicular to the crystallographic $c$-axis.  Upper inset: enlarged transition region. Lower inset: anisotropic magnetization for applied field parallel and perpendicular to the $c$-axis for $T = 300$ K. As described in text, the susceptibility was determined by the difference between 30 kOe and 50 kOe magnetization runs so as to eliminate weak ferromagnetic contributions.}\label{F5}
\end{figure}

Magnetization data on single crystal CaFe$_2$As$_2$ are presented  in Fig. \ref{F5}. Magnetic-field-dependent, anisotropic magnetization, taken at 300 K (lower inset to Fig \ref{F5}) suggest the possible presence of a small ferromagnetic impurity, or, less likely, an intrinsic, weak ferromagnetic transition above 300 K. In attempt to eliminate the effect of this weak ferromagnetism, the temperature dependent, anisotropic, susceptibility data are presented as $(M_{50kOe}(T) - M_{30kOe}(T))/(50kOe-30kOe)$ in Fig. \ref{F5}.  The susceptibility of CaFe$_2$As$_2$ is weakly anisotropic with neither direction of applied field showing any sign of local moment-like temperature dependence.  The clearest feature in the data set is the sharp increase in magnetization seen when the sample is warmed through 171 K. The change in susceptibility at the transition is $3 - 4$ times larger for $H \| ab$.  The onset of the transition at 171 K for data collected upon warming is consistent with the hysteresis of the transition detected in the resistivity data shown in Fig. \ref{F3} above.

\begin{figure}
\begin{center}
\includegraphics[angle=0,width=120mm]{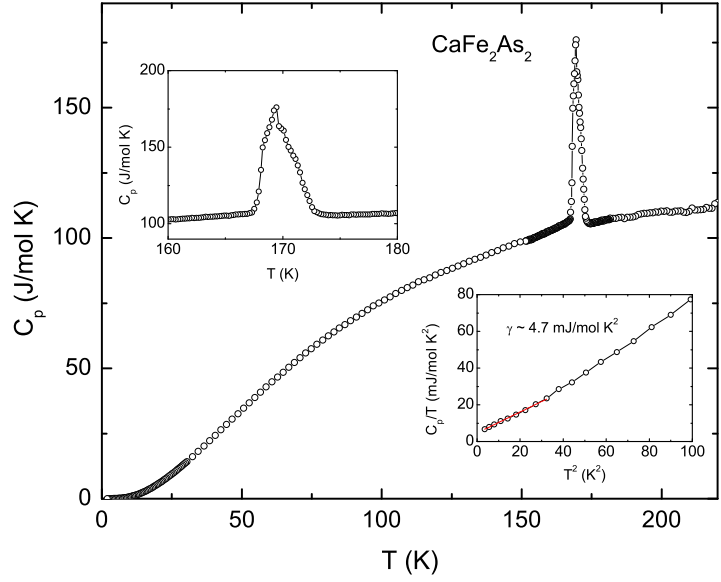}
\end{center}
\caption{Temperature dependent specific heat of CaFe$_2$As$_2$.  Upper inset:  enlargement of data in the vicinity of the 170 K phase transition.  Lower inset:  $C_p/T$ plotted as a function of $T^2$; line - linear fit.}\label{F6}
\end{figure}

Figure \ref{F6} presents the temperature dependent specific heat of CaFe$_2$As$_2$.  There is a sharp anomaly centered on 170 K in the cooling data.  This is shown more clearly in the upper inset.  The data were collected with a thermal excitation of 2 \% of the temperature (i.e. $\sim 4$ K) in the temperature range of the anomaly, causing the apparent broadening of the latent heat feature.  The low temperature $C_p/T$ data are plotted as a function of $T^2$ in the lower inset.  These data can be fit to a $C_p(T) = \gamma T + \beta T^3$ power law giving $\gamma \approx 4.7$ mJ/mol K$^2$ and $\beta = 0.56$ mJ/mol K$^4$  ($\Theta_D \approx 258$ K).  Whereas the $\beta$ ($\Theta_D$) values are similar to those seen for BaFe$_2$As$_2$ and SrFe$_2$As$_2$, it is worth noting that the $\gamma$ value is reduced by almost an order of magnitude.

\begin{figure}
\begin{center}
\includegraphics[angle=0,width=120mm]{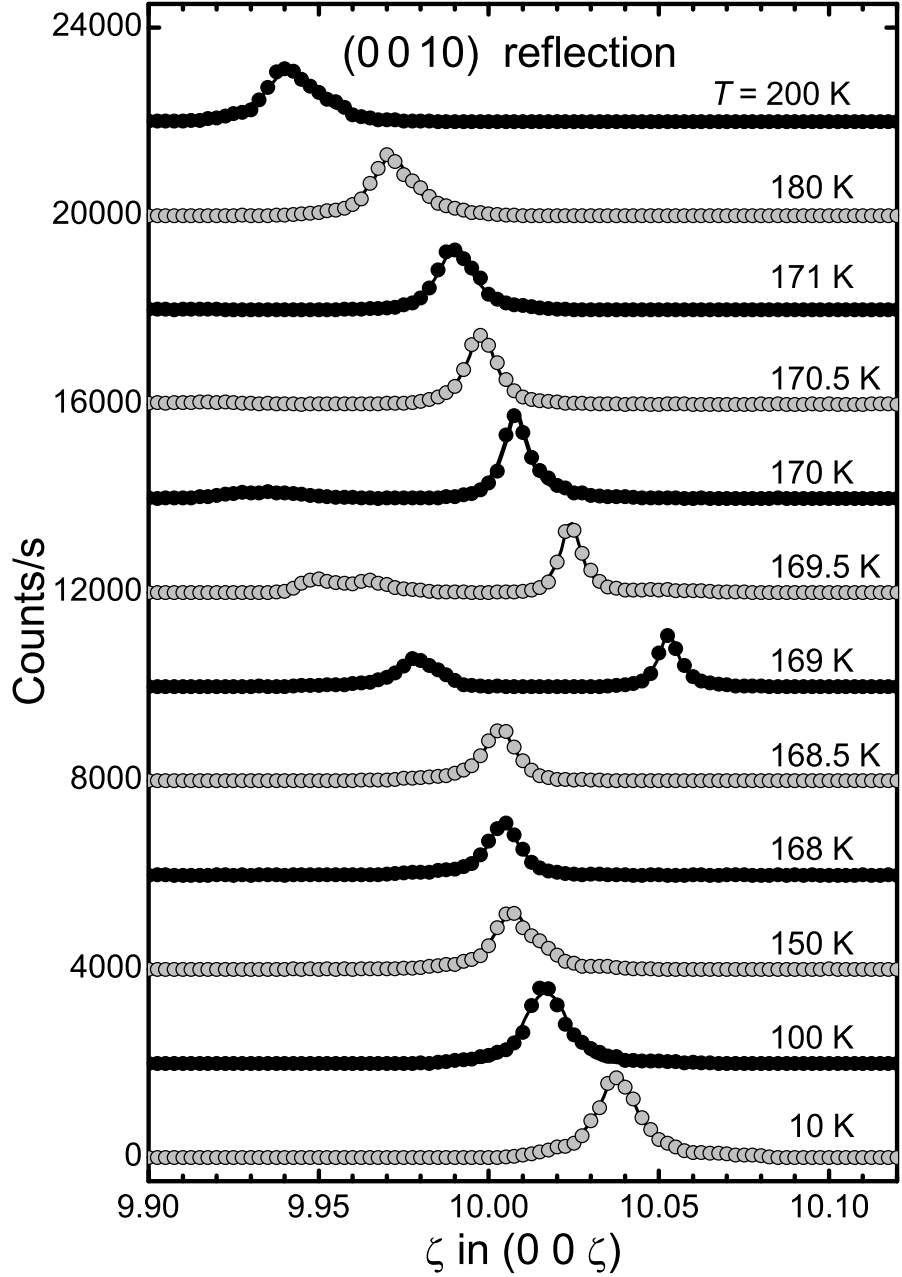}
\end{center}
\caption{Longitudinal $(0~0~\xi)$ scans through the position of the tetragonal $(0~0~10)$ reflection for selected temperatures. The lines represent fits to the data to obtain the reflection positions and intensities shown in Figure \ref{F9}. The offset between every data set is 2000 Counts/s. In the coexistence range, we point out that the strong change in the position of the peaks results mainly from misalignment related to the strain occurring at the phase transition in combination with the lattice-parameter changes between both phases.}\label{F7}
\end{figure}

\begin{figure}
\begin{center}
\includegraphics[angle=0,width=120mm]{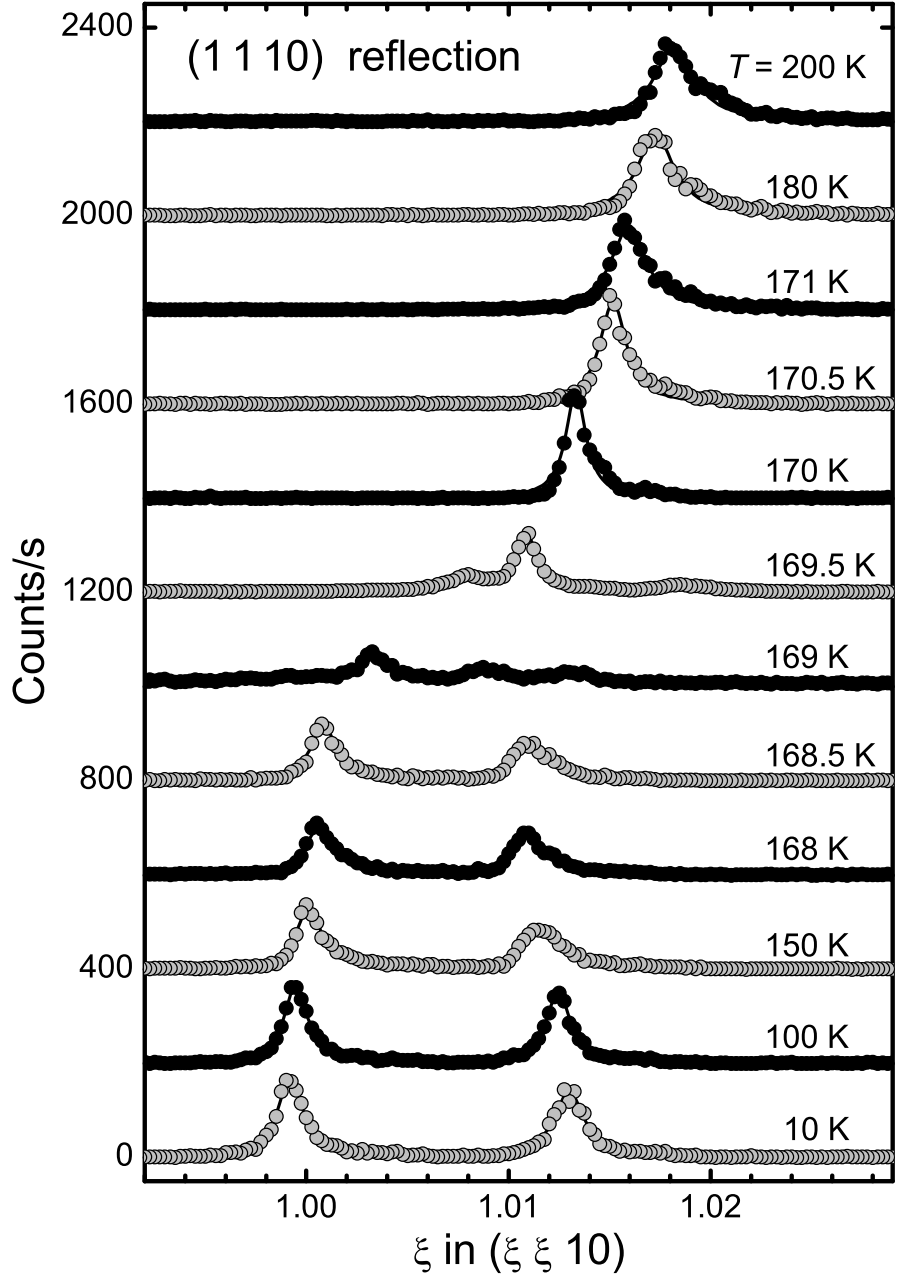}
\end{center}
\caption{Transverse $(\xi~\xi~0)$ scans through the position of the tetragonal $(1~1~10)$ reflection for selected temperatures. The lines represents to the data to obtain the reflection positions shown in Figure \ref{F9}. The offset between every data set is 200 Counts/s. In the coexistence range, we point out that the strong change in the position of the peaks results mainly from misalignment related to the strain occurring at the phase transition in combination with the lattice-parameter changes between both phases. }\label{F8}
\end{figure}

The thermodynamic and transport data shown in Figs. \ref{F3},\ref{F5},\ref{F6} are consistent with a first order phase transition near 170 K.  To determine whether there is a structural transition, as observed in the isostructural AFe$_2$As$_2$ (A = Ba, Sr) compounds \cite{5,8,9}, we performed a single crystal x-ray diffraction study.  $(\xi~\xi~0)$ and $(0~0~\xi)$ scans through the $(0~0~10)$ reflection show no change in the shape of the $(0~0~10)$ reflection. However, in a narrow temperature range between 168.5 K and 171 K we observe coexistence of two reflections in longitudinal $(0~0~\xi)$ scans as illustrated in Fig. \ref{F7}. The reflection at lower $\xi$ values in the $(0~0~\xi)$ scans corresponds to the $(0~0~10)$ reflection of the low temperature phase. The reflection at higher $\xi$ values represents the high-temperature phase. In transverse $(\xi~\xi~0)$ scans, however, only one sharp reflection was observed. The abrupt change in the $c$-lattice parameter, together with the narrow temperature range of coexistence of both phases, indicates a first order phase transition at $T \sim 170$ K . This interpretation is corroborated by our measurements of the $(1~1~10)$ reflection. As shown in Fig. \ref{F8}, below 170 K, a splitting of the $(1~1~10)$ reflection was observed in $(\xi~\xi~0)$ scans, consistent with a tetragonal-to-orthorhombic phase transition with a distortion along the diagonal $(1~1~10)$ direction. This transition from the space group $I4/mmm$ to $Fmmm$ is similar to that observed in the AFe$_2$As$_2$ (A = Ba, Sr) compounds. \cite{5,8,9} Between 168.5 K and 171 K, reflections related to both phases coexist. The central peak corresponds to the $(1~1~10)$ reflection of the tetragonal high-temperature phase. The surrounding pair of reflections arises from the two sets of 'twin domains' that arise from the orthorhombic distortion along the diagonal $(1~1~0)$ direction for the low-temperature phase.

\begin{figure}
\begin{center}
\includegraphics[angle=0,width=100mm]{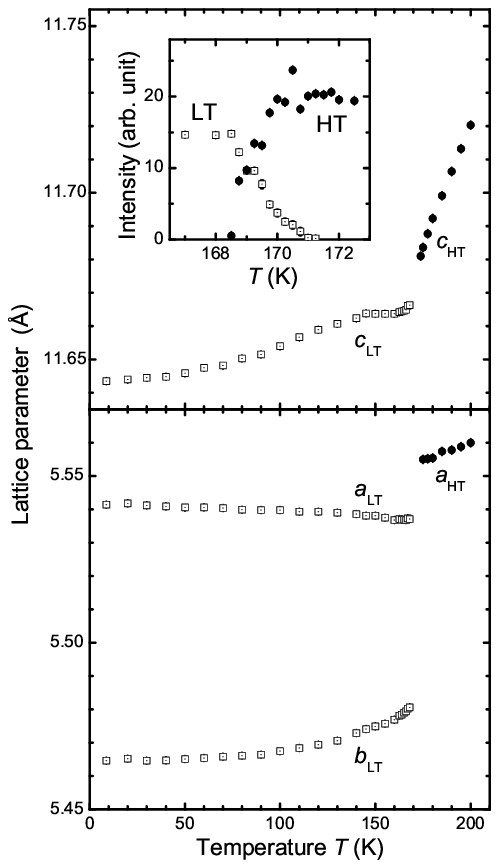}
\end{center}
\caption{Lattice parameters for the tetragonal and orthorhombic phases as extracted from the data shown in Figures \ref{F7} and \ref{F8} for the $(0~0~10)$ reflection and the $(1~1~10)$ reflection. To allow direct comparison of both phases the lattice parameter $a$ of the tetragonal phase is given as $a_{\textrm{HT}}=\sqrt{2}a$. The inset shows the intensity of the reflections related to the lattice parameter $c$ (shown in Fig. \ref{F7} in both phases). The error bars represent only the relative precision of the measurements.}\label{F9}
\end{figure}

By analyzing the position of the $(0~0~10)$ reflection in longitudinal $(0~0~\xi)$ scans, the $c$-lattice parameter can be determined. The in-plane $a$ and $b$-lattice parameters have been calculated based on the distance between the reflections close to the tetragonal $(1~1~10)$ position and the $(0~0~10)$ reflection in transverse scans along the $(\xi~\xi~0)$ direction. The results are shown in Fig. \ref{F9}.  Between 10 K and 170 K, the difference in the orthorhombic $a$ and $b$-lattice parameters decreases slowly and monotonically with increasing temperature whereas the $c$-lattice parameter increases monotonically up to 170 K. At 170 K, the $a$-lattice parameter jumps abruptly as can be readily seen from the transverse $(\xi~\xi~0)$ scans through the $(1~1~10)$ reflection in Fig. \ref{F8} and the temperature dependence of lattice parameters in Fig. \ref{F9}. The jump is as large as $\sim 1$\% between the high-temperature tetragonal value and the averaged low-temperature in-plane lattice parameters $a$ and $b$.  This change is quite large and emphasizes the first order nature of the structural phase transition at $T_0 = 170$ K.  The $c$-lattice parameter shows also a jump at $T_0$, but smaller than the in-plane lattice parameter.

The inset to Fig. \ref{F9} shows the intensity of the low- and high-temperature $(0~0~10)$ reflection  as a function of temperature close to the tetragonal to orthorhombic phase transition. The overall behavior reflects again the first order character of the phase transition as well as the high quality of the sample. Coexistence of the high-temperature tetragonal and low-temperature orthorhombic phase can be detected for a very narrow temperature range of only $\sim 2$ K representing the 'sharpness' of the transition in the studied sample.

\section{Discussion and Summary}
CaFe$_2$As$_2$ single crystals have been grown from Sn flux and have been found to form in the tetragonal, ThCr$_2$Si$_2$ structure that is the high temperature structure of the related BaFe$_2$As$_2$ and SrFe$_2$As$_2$ compounds.  Unlike single crystals of BaFe$_2$As$_2$ and SrFe$_2$As$_2$ grown from Sn flux, CaFe$_2$As$_2$ single crystals have no detectable Sn incorporated into the bulk.

CaFe$_2$As$_2$ manifests a clear, first order structural phase transition near 170 K in thermodynamic, transport and microscopic data.  The phase transition is from a high temperature tetragonal, to a low temperature orthorhombic phase.  This phase transition is similar to that seen in BaFe$_2$As$_2$ and SrFe$_2$As$_2$, compounds that manifest superconductivity up to near 40 K when K is substituted for Ba or Sr.  In CaFe$_2$As$_2$ this structural phase transition is unambiguously first order in nature, showing discontinuous jumps in electrical resistivity (as well as hysteresis), 
magnetic susceptibility and lattice parameters.  In addition, the specific heat data show a sharp spike-like feature near 170 K that has a width comparable to the thermal excitation used in the measurements.
	These results raise the possibility that, if K can be substituted for Ca, or some other form of hole doping can be achieved, then doped CaFe$_2$As$_2$ could also manifest a suppression of this structural phase transition and become a third member of this family to manifest superconductivity.

\begin{acknowledgments}
Work at the Ames Laboratory was supported by the US Department of Energy - Basic Energy Sciences under Contract No. DE-AC02-07CH11358.  We would like to acknowledge useful discussions and assistance from, F. Laabs, J. Q. Yan, R. W. McCallum, M. Lampe, and S. Moser.

\end{acknowledgments}

\clearpage

\clearpage

\clearpage

\clearpage

\clearpage

\clearpage

\clearpage

\clearpage

\begin{thebibliography}{99}

\bibitem{1}  Y. Kamihara, T. Watanabe, M. Hirano, H Hosono, Journal of the American Chemical Society  {\bf 130}, 3296 (2008).

\bibitem{2}  H. Takahashi, K. Igawa, K. Arii, Y. Kamihara, M. Hirano, H. Hosono, Nature (London)  {\bf 453}, 376 (2008).

\bibitem{3}  X. H. Chen, T. Wu, G. Wu, R. H. Liu, H. Chen, D. F. Fang, Nature (London)  {\bf 453}, 761 (2008).

\bibitem{4}  M. Rotter, M. Tegel, D. Johrendt, arXiv:0805.4630, unpublished.

\bibitem{5} N. Ni, S. L. Bud'ko, A. Kreyssig, S. Nandi, G. E. Rustan, A. I. Goldman, S. Gupta, J. D. Corbett, A. Kracher, P. C. Canfield, arXiv:0806.1874, unpublished.

\bibitem{6} G. F. Chen, Z. Li, J. Dong, G. Li, W. Z. Hu, X. D. Zhang, X. H. Song, P. Zheng, N. L. Wang, and J. L. Luo, arXiv:0806.2648, unpublished.

\bibitem{7} X. F. Wang, T. Wu, G. Wu, H. Chen, Y. L. Xie, J. J. Ying, Y. J. Yan, R. H. Liu, X. H. Chen, arXiv:0806.2452, unpublished.

\bibitem{8} J.-Q. Yan, A. Kreyssig, S. Nandi, N. Ni, S. L. Bud'ko, A. Kracher, R. J. McQueeney, R. W. McCallum, T. A. Lograsso, A. I. Goldman,
P. C. Canfield, arXiv:0806.2711, unpublished.

\bibitem{9}  M. Rotter, M. Tegel, D. Johrendt, I. Schellenberg, W. Hermes, R. Poettgen, arXiv:0805.4021, unpublished.

\bibitem{ad} C. Krellner, N. Caroca-Canales, A. Jesche, H. Rosner, A. Ormeci, and C. Geibel, arXiv:0805.1043, unpublished.

\bibitem{10}  P. C. Canfield, Z.  Fisk, Phil. Mag. B  {\bf 65}, 1117 (1992).

\end{thebibliography}
\end{document}